\documentclass[journal]{vgtc}                % final (journal style)
\ifpdf%                                % if we use pdflatex
  \pdfoutput=1\relax                   % create PDFs from pdfLaTeX
  \usepackage{graphicx}                % allow us to embed graphics files
  \DeclareGraphicsExtensions{.pdf,.png,.jpg,.jpeg} % for pdflatex we expect .pdf, .png, or .jpg files
\else%                                 % else we use pure latex
  \usepackage{graphicx}                % allow us to embed graphics files
  \DeclareGraphicsExtensions{.eps}     % for pure latex we expect eps files
\fi%

\graphicspath{{figures/}{pictures/}{images/}{./}} % where to search for the images

\usepackage{microtype}                 % use micro-typography (slightly more compact, better to read)
\PassOptionsToPackage{warn}{textcomp}  % to address font issues with \textrightarrow
\usepackage{textcomp}                  % use better special symbols
\usepackage{mathptmx}                  % use matching math font
\usepackage{times}                     % we use Times as the main font
\usepackage{cite}                      % needed to automatically sort the references
\usepackage{tabu}                      % only used for the table example
\usepackage{booktabs}                  % only used for the table example
\usepackage{subfiles}
\usepackage{kotex}
\usepackage{enumitem}
\usepackage{subfigure}
\usepackage{array,multirow,graphicx}
\usepackage{float}

\usepackage{amsmath}
\usepackage{algorithm}
\usepackage[noend]{algpseudocode}

\usepackage{tabularx}

\makeatletter
\def\BState{\State\hskip-\ALG@thistlm}
\makeatother

\newcommand{\parahead}[1]{{\vspace{0.05cm}\fontfamily{phv}\selectfont{#1} \ \ }}
\onlineid{1117}

\vgtccategory{Research}
\vgtcpapertype{application/design study}

\title{Githru: Visual Analytics for Understanding Software Development History Through Git Metadata Analysis}

\author{
Youngtaek Kim, Jaeyoung Kim, Hyeon Jeon, Young-Ho Kim, Hyunjoo Song, Bohyoung Kim, and Jinwook Seo
}
\authorfooter{
\item
 Youngtaek Kim, Jaeyoung Kim, and Jinwook Seo are with Seoul National University. E-mail: \{ytaek.kim, jykim\}@hcil.snu.ac.kr, jseo@snu.ac.kr.
\item
 Youngtaek Kim is also with Samsung Electronics.
\item
 Hyeon Jeon is with POSTECH. E-mail: jeonhyun97@postech.ac.kr.
\item
 Young-Ho Kim is with University of Maryland. E-mail: yghokim@umd.edu.
 \item
 Hyunjoo Song is with Soongsil University. E-mail: hsong@ssu.ac.kr.
\item
 Bohyoung Kim is with Hankuk University of Foreign Studies. \newline
 E-mail: bkim@hufs.ac.kr.
}

\shortauthortitle{Biv \MakeLowercase{\textit{et al.}}: Global Illumination for Fun and Profit}
\abstract{
Git metadata contains rich information for developers to understand the overall context of a large software development project. Thus it can help new developers, managers, and testers understand the history of development without needing to dig into a large pile of unfamiliar source code.
However, the current tools for Git visualization are not adequate to analyze and explore the metadata: They focus mainly on improving the usability of Git commands instead of on helping users understand the development history. Furthermore, they do not scale for large and complex Git commit graphs, which can play an important role in understanding the overall development history.
In this paper, we present Githru, an interactive visual analytics system that enables developers to effectively understand the context of development history through the interactive exploration of Git metadata. We design an interactive visual encoding idiom to represent a large Git graph in a scalable manner while preserving the topological structures in the Git graph.
To enable scalable exploration of a large Git commit graph, we propose novel techniques (graph reconstruction, clustering, and Context-Preserving Squash Merge (CSM) methods) to abstract a large-scale Git commit graph. Based on these Git commit graph abstraction techniques, Githru provides an interactive summary view to help users gain an overview of the development history and a comparison view in which users can compare different clusters of commits.
The efficacy of Githru has been demonstrated by case studies with domain experts using real-world, in-house datasets from a large software development team at a major international IT company.
A controlled user study with 12 developers comparing Githru to previous tools also confirms the effectiveness of Githru in terms of task completion time.
} % end of abstract


\CCScatlist{ % not used in journal version
 \CCScat{K.6.1}{Management of Computing and Information Systems}%
{Project and People Management}{Life Cycle};
 \CCScat{K.7.m}{The Computing Profession}{Miscellaneous}{Ethics}
}

\teaser{
  \makebox[\textwidth][c]{\includegraphics[height=0.33\paperheight]{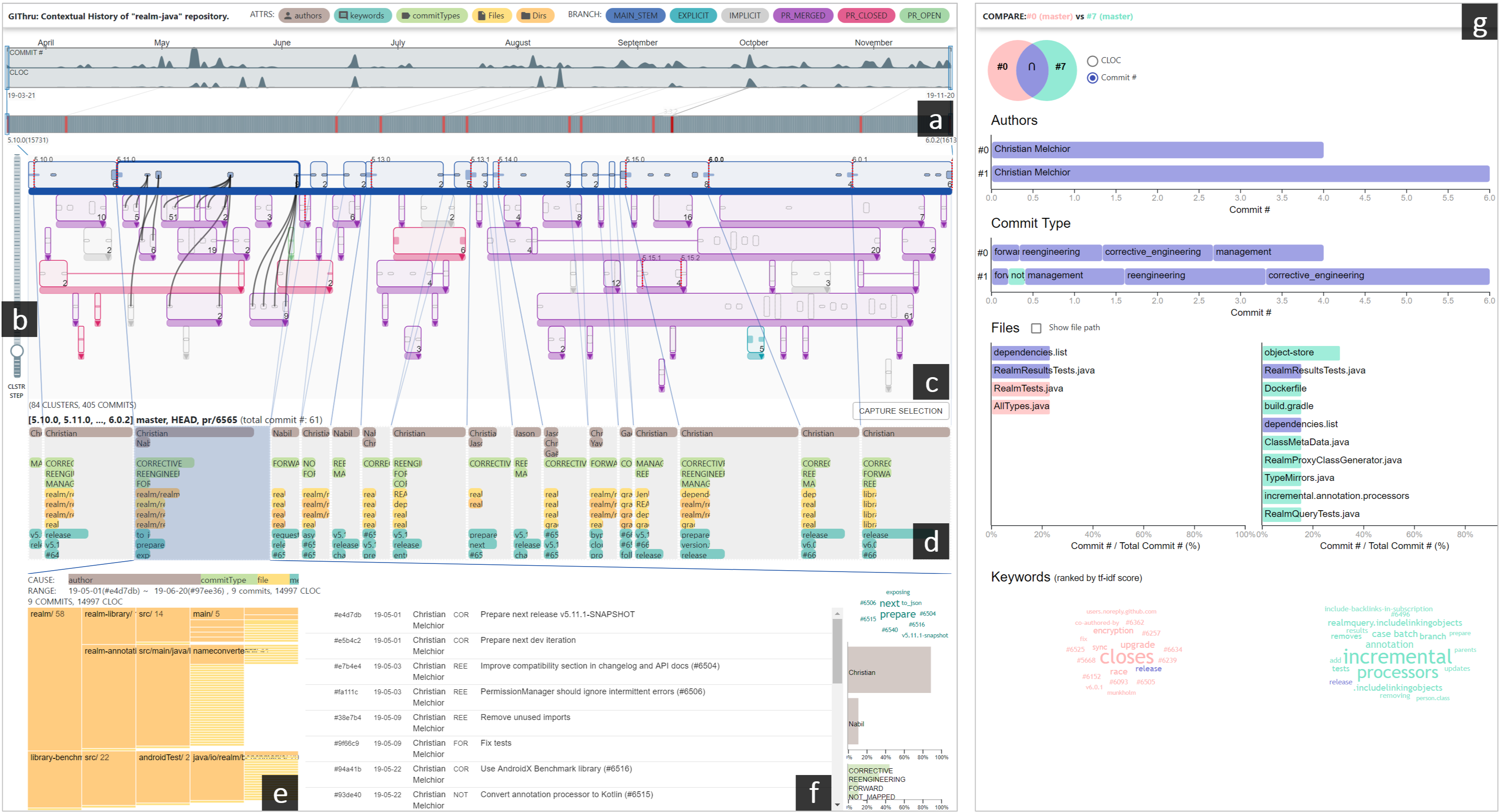}}%
  \vspace{-0.05cm}
  \caption{
    Githru system.
    (a) The Global Temporal Filter shows commit trends by number of commits and CLOC (changed lines of code).
    (b) The Clustering Step controls the granularity of clustering.
    (c) The \emph{stem} graph visualizes a cluster information of each commit at a single glance.
    (d) The Grouped Summary View provides a rough summary of the selected clusters.
    (e) A file icicle tree allows users to interactively observe the modified file hierarchy.
    (f) The commit list shows all the commits in a selected cluster.
    (g) Comparison View enables a comparison between selected groups.
  }
  \label{fig:teaser}
}

\vgtcinsertpkg

\begin{document}


\firstsection{Introduction}
\maketitle
%% \section{Introduction} %for journal use above \firstsection{..} instead
The Git repository archives the development history of a project.
It keeps a record of the changes, called “commits”, made to a file or set of files, along with metadata for each commit such as the unique ID, author, message, and date.
It also maintains information about branches (or merges) created to diverge from (or converge to) the main line of development.
Thus, the metadata of the Git repository is known to provide rich information for understanding the overall context of development history without delving directly into the source code \cite{north2016understanding}.
As new developers, managers, and testers of a project are generally not well acquainted with the source code, such metadata could alleviate the burden of comprehending the code themselves.

However, as the size of a project grows in terms of the number of commits, branches, authors, or files, going through a long list of commit logs becomes challenging.
Several Git clients show the graphical representation of a repository as a directed acyclic graph (DAG), where each node is a commit and each edge represents the parent-child relationship between commits \cite{elsen2013visgi}. 
While such visualizations can alleviate the scalability issue to some extent, most Git clients focus on improving the usability of Git commands rather than on helping users understand the development history.
Some prior studies have provided a graphical overview and supported visual exploration of development history, but still lack visual analytics functions that provide all the metadata while preserving the branch and merge information of the Git graph.
Moreover, the visualization scalability issue with large DAGs remains unresolved in such studies.

In this paper, we introduce Githru, an interactive visual analytics system for Git metadata, to help users explore and understand the context of development history.
We propose novel analytic techniques to deal with the complexity and scalability issues of large sets of Git metadata. 
We start by identifying and removing the redundant nodes in the DAG that come from a conventional git merge operation.
Such a reconstruction simplifies the topology of a Git commit graph, which greatly reduces the cognitive load of branch traversal.
We also introduced additional techniques to simplify the topology by grouping similar nodes according to user-defined criteria.
Lastly, based on the Git commit graph abstraction techniques, Githru provides an interactive summary view to help users gain an overview of the development history and a comparison view to enable users to compare different clusters of interest.
We followed an iterative design process to design Githru, closely collaborating with developers and managers at Samsung Electronics to formulate requirements; created and refined design prototypes; and evaluated the system.
To assess the efficacy and usability of Githru, we performed case studies with an in-house repository dataset, followed by interviews with domain experts.
We also conducted a controlled user study with 12 developers comparing Githru to a combination of existing widely used tools: GitHub and git log.

The major contributions of this paper are as follows:
\setlist{nolistsep}
\begin{itemize}[noitemsep]
    \item We define, through literature reviews and expert interviews, a list of tasks to analyze large sets of Git metadata.
    \item We propose a set of new analytics techniques to abstract large and complex Git commit graphs. 
    \item We present novel visualization designs for the interactive exploration of large-scale Git metadata.
    \item We evaluate Githru with domain experts using real-world, in-house datasets.
\end{itemize}

\section{Background}

In this section, we introduce the background of Git metadata. We also identify target users to analyze development history through preliminary interviews.

\subsection{Git Metadata}
Git metadata are a collection of content for each commit, such as its unique ID, author, message, date, and information on the set of changed files.
A commit is connected to its parent commits, and thus the metadata can be represented as a DAG, where each node is a commit and each edge is the parental relationship between the commits \cite{bird2009promises}. The topology of a DAG encodes the history of the activity of commits. Hence, users need to explore a DAG to understand the overall development history.

The complexity of a DAG with Git metadata mainly depends on the edges created through branching in Git.
Branching refers to diverging from the main line of development and continuing to work without affecting that main line.
In Git, a branch is simply a lightweight movable pointer to a commit, and therefore, branches are cheap to create and destroy \cite{10.5555/2695634}. Due to this low cost, Git encourages a workflow that branches and merges often.

\begin{figure}[t]
    \centering
    \includegraphics[width=\linewidth]{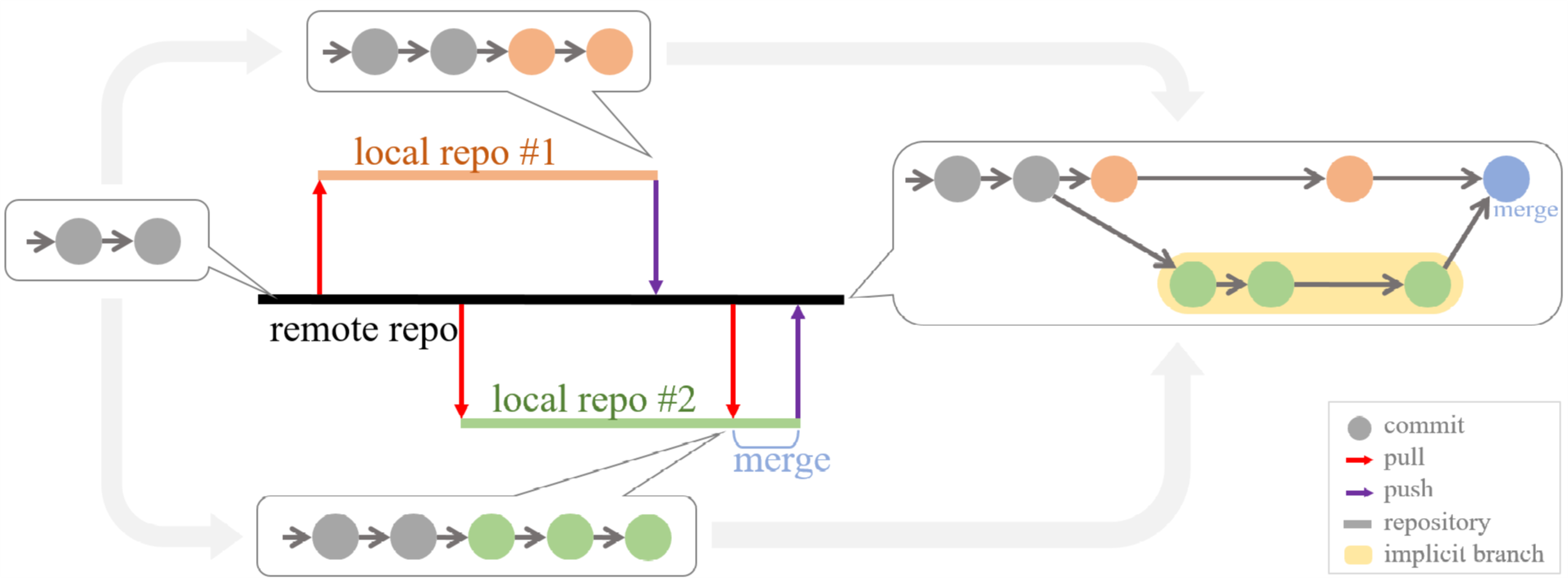}
    \caption{An unintended implicit branch is created when synchronized with the remote repository.}
    \label{fig:automerge}
\end{figure}

At the same time as an unintended side effect of decentralization, Git can also create implicit branches as illustrated in  Fig. \ref{fig:automerge} \cite{bird2009promises, Kovalenko2018}. 
Suppose that a developer pulled the latest snapshot from the remote repository, added several commits to the local repository (repo \#1), and pushed them back. 
If another developer (with repo \#2) then tried to push additional commits to the remote repository, the synchronization with the remote repository could introduce an implicit branch.

Due to the frequent branching and unintended implicit branches, the number of branches increases drastically as the project grows.
Therefore, the complexity of the DAG increases, which complicates the interpretation of the graph to understand the overall development history.
Such high complexity also aggravates the cognitive load of traversing parent commits when identifying which commits are finally merged into the main line.

If a Git repository is hosted on GitHub, pull requests (PRs) can provide additional information.
Pull requests are proposed changes to a repository. Submitted PRs by users are accepted or rejected by the repository's maintainers \cite{kalliamvakou2014promises}. 
A PR operates identically to an explicit branch in the graph but provides additional information, such as a message written by the creator of the PR, the state (open, closed, or merged), and the commit into which the PR is merged. 

In this study, we extract the metadata from the results of the \texttt{git log} command and augment them with additional information. First, we add a commit type that represents the characteristics of a commit in accordance with developers' needs as derived from the preliminary interviews. A commit type is classified using Hattori's method \cite{hattori2008nature} as a baseline.
Next, we add the number of added and deleted lines of code (LOCs) per each changed file provided by \texttt{git log} with the \texttt{--numstat} option \cite{gitLogDocumentation}. We also use the information in PRs from the GitHub data.

\subsection{Target Users}
The preliminary interviews were conducted in a semi-structured form to ascertain the practical necessity of a metadata analysis and to identify the types of users who need to analyze development history.
The interviewees were five software engineers: three developers (D1--3) and two project leaders (L1--2) who perform both management and development. Their development experience ranged from 5 to 16 years (mean, 10.2 years), and they had worked in engineering, infrastructure, and AI.

All interviewees confirmed the necessity of understanding the development history.
As for metadata analysis, L1 and L3 said that they did not try analysis due to a lack of time.
The others said that they felt performing analyses with the existing tools, such as \texttt{git log} in a console, Sourcetree \cite{sourcetree}, or the GitHub website, was inconvenient. Normally they navigated a long list of commits to obtain an overview of the commit history and summarized what they wanted to focus on.
They sometimes wrote scripts to answer questions.

During the interviews, we discussed the types of target users who need to analyze development history in practical terms.
Developers who are contributing to the source code of ongoing projects feel less of a need to explore and gain an overview of historical data with visual tools: They are already acquainted with the context, so they know what or where the answer is.
Hence, in this study, we focus on new developers, managers, and testers who are not yet familiar with the source codes as the target users.

A new developer who has just joined a new team needs to fully understand the previous works and source codes of the latest snapshot.
The information can be obtained from the previous developer, project documents, or the infrastructure system. When such necessary information is missing, a Git metadata analysis can answer questions such as, Which component was John mainly responsible for recently? Who were John's co-workers? and How much has he been working since the last quarter? \cite{fritz2010using}.

Similarly, managers and testers also need to deal with historical information.
Managers need an overview of individual developers' work. They often explore areas where many changes have occurred and figures out what happened at that point. 
Testers also need to understand the development history to write test cases \cite{tao2012software} or perform bug localization \cite{sisman2012incorporating} and triage \cite{hu2014effective}.

\section{Related Work} 

The design of Githru is inspired by numerous prior studies. In this section, we introduce the core research areas related to Githru: software evolution visualization, Git metadata analysis, and historical questions.

\subsection{Software Evolution Visualization}

Stephan Diehl categorized software visualization into three aspects: \emph{structure}, \emph{behavior}, and \emph{evolution} \cite{diehl2007software}. As Githru mainly focuses on visualizing development history, it can be classified as \emph{evolution}.
We first investigate the existing tools that visualize software evolution \cite{alcocer2019performance, wang2019clonecompass, alexandru2019evo, perrie2019city, porkolab2018codecompass, lebeuf2018understanding, ulan2018quality, feist2016visualizing, tymchuk2016walls, burch2015visualizing, khan2015interactive, rufiange2014animatrix, ens2014chronotwigger, yoon2013visualization, anslow2013sourcevis}.
SeeSoft \cite{eick1992seesoft} offers a line-oriented visualization of source code, mapping the software attributes to different line colors. Through this mapping, developers can see who worked on which specific lines of code. Similarly, Augur \cite{froehlich2004unifying} and Advizor \cite{eick2002visualizing} visualize developer-related data. These systems provide a quantitative way to analyze software. 

GitHub \cite{github} features various built-in visualizations to support exploration in \emph{GitHub Insights}. However, they are still in the early stages and unable to effectively support the analytical tasks users would like to perform. For example, GitHub provides a graph showing LOC changes over time but does not support chronological filtering. Also, users cannot look at the commit graphs at a glance and must scroll significantly. 
RepoVis \cite{feiner2018repovis} provides a comprehensive visual overview and full-text search for Git repositories, which enables a rapid overview of the repository. However, it shows only a snapshot of a revision, not the overall history. 
ConceptCloud \cite{greene2014conceptcloud} is a web application that automatically indexes Git and SVN repositories and summarizes them through interactive tag clouds. 
Another novel approach to software evolution visualization is Code Swarm \cite{ogawa2009code_swarm}, which applies animation and organic information visualization techniques to create aesthetic visualizations of software projects for casual viewers. 

To help understand Git DAGs, git log \cite{gitLogDocumentation} provides a graphical representation in a console environment. 
Many Git GUI clients, such as Sourcetree \cite{sourcetree}, GitKraken \cite{gitkraken}, gitk \cite{gitk} and gmaster \cite{gmaster}, represent DAGs similarly to git log. 
On the other hand, Elsen has introduced VisGi \cite{elsen2013visgi}, which compresses complex Git repository structures into DAGs.

% reconstruction of DAG structure
Various reconstruction methods of DAGs have been proposed to solve the problem of high complexity. Wilde proposed Linvis \cite{wilde2018merge}, a web-based tool that uses the conversion of a DAG to merge trees. Linvis also proposed visualization to represent the hierarchy of merged trees. Gitgraph \cite{zhao2019knowledge} constructs a knowledge graph that users can interactively explore and that helps developers comprehend software repositories. However, when a DAG is reconstructed, the original topological information is lost or distorted. In contrast, Githru preserves the chronological sequence of commits with reduced complexity.

\subsection{Git Metadata Analysis}
Git repository data have been actively studied to understand various facets of development history. Bird and Kalliamvakou provided a set of recommendations for researchers on how to approach Git and GitHub data \cite{bird2009promises, kalliamvakou2014promises}. 
Stevens \cite{stevens2014querying} presented the QWALKEKO meta-programming library, which enables querying the history of versioned software projects in a declarative manner. Barik \cite{barik2015commit} proposed Commit Bubbles, which supports developers in constructing commit histories with their coding workflows. However, Commit Bubbles is merely conceptual and has not been implemented. Rozenberg introduced RepoGram \cite{rozenberg2016comparing}, a tool to support comparing and contrasting tasks in software projects over time.

Branching is one of the main features of Git. Various tools have been developed to support the analysis of its topology. 
Lee et al. \cite{lee2013git} proposed a tool that extracts branch data from Git repositories and abstracts each commit and branch into the workflow.
Biazzini et al. \cite{biazzini2014analyzing} defined Metagraph, a data structure representing topologically relevant commits. 
Based on the definition, they analyzed the topological characteristics of Git repositories and identified patterns recurring in multiple repositories.

Githru also covers historical aspects of development using Git metadata; however, it focuses on the main branch and reduces complexity of visualization by simplifying other branches to resolve scalability issues. Section 5 provides a detailed explanation.

\subsection{Historical Questions}
There have been many studies on the questions that should be answered to augment the understanding of development history. After interviewing 203 participants, Begel and Zimmermann \cite{begel2014analyze} presented a ranked list of 145 questions that software engineers want data scientists to answer. Fritz and Murphy \cite{fritz2010using} and Buse et al. \cite{buse2012information} also identified questions that developers have about projects. LaToza et al. \cite{latoza2010hard} organized 94 distinct questions about code that are hard to answer. All three studies were conducted by interviewing professional developers. 
Silito et al. \cite{sillito2006questions} categorized 44 different kinds of questions about the information programmers need and how they discover it. Kubelka et al. \cite{kubelka2019live} conducted software evolution sessions in Live Programming and assigned questions to each session, based on Silito et al.'s study. Codoban et al. \cite{codoban2015software} performed an empirical study about the motivations developers have for examining software history, the strategies they use, and the challenges they encounter. 

Research also exists about how developers analyze their code and repositories. 
Sadowski et al. \cite{sadowski2015developers} analyzed how developers search for code and provided insights into multiple aspects, including what developers are doing and trying to learn when performing a search. Tao et al. \cite{tao2012software} explored the information engineers' need to understand changes and their requirements for the corresponding tool. Safwan and Servant \cite{safwan2019decomposing} discovered how developers decompose the rationale for code commits in the context of software maintenance. Interviewing 20 software developers allowed these authors to understand their experience. 

The design of Githru is based on the analytics tasks extracted and organized from previous studies. Section 4.1 describes these in detail.

\section{Requirement Analysis}
We extracted analytics questions from our literature review and validated them through interviews with five experts. We then organized them into three analytics tasks and formulated requirements accordingly based on the tasks. 
 
\subsection{Task Abstraction}

To construct analytics tasks, we investigated 1,479 papers from software engineering conferences, including ICSE (660), FSE (574), and ICSME (245), by querying “question AND (history OR evolution)” for papers since 2010. We selected four papers \cite{safwan2019decomposing, tao2012software, fritz2010using, sadowski2015developers} suggesting at least five history-related questions that can be answered by Git metadata without analyzing source code. Two additional papers \cite{hattori2011software, latoza2010hard} were included subsequently by investigating the citations of the selected papers under the same criteria. We derived nine exemplary questions by analyzing the six papers using thematic analysis \cite{braun2006using}. 

To validate whether the questions reflect real-world problems, we conducted semi-structured interviews with five software engineers (two developers and three project leaders, D2--3 and L1--3). They had an average of 11.8 years (ranging from 7 to 16 years) of professional software development experience and had worked in the areas of engineering, infrastructure, AI, and cloud computing.
We asked the interviewees how the questions relate to real-world problems. Through the interviews, we formulated three significant tasks as follows:

\setlist{topsep=0.1em}
\begin{itemize}[itemsep=1pt, leftmargin=1.5em]
% \begin{itemize}[label={}, leftmargin=*]
% \setlist{nolistsep}
% \begin{itemize}[noitemsep, leftmargin=*]
    \item \textbf{T1: Understand the overall development context.}  
    Obtain an overview of the development history in terms of mainly temporal aspects.
    \textit{Who is working on what?} \textit{What are my co-workers working on right now?} and \textit{What has changed between two builds?} \cite{fritz2010using}
    
    \item \textbf{T2: Understand the topology of commit history.}
    Analyze the merging and branching.
    \textit{Have changes in another branch been integrated into this branch?} and \textit{Is the pull request “501” merged into the master in this release?} \cite{latoza2010hard}
    
    \item \textbf{T3: Explore and compare details.}
    Interactively explore detailed information (e.g., directory containing changed files, name of author, keywords of message) that meets user-defined criteria; and compare the context between multiple periods to find meaningful patterns (e.g., hotspots \cite{tao2012software} where frequent or large numbers of changes have occurred).
    \textit{Which component was John mainly responsible for recently?} \textit{Is this changed location a hotspot for past changes?} and \textit{How many LOC have been changed?} \cite{safwan2019decomposing, tao2012software}
\end{itemize}

%%%% TABLE

% \renewcommand{\cs}{0.11799}
\newcommand{\cs}{0.11799}
\newcommand{\sistretch}{0.52mm}
\newcommand{\bst}{\hspace*{0.42mm}}

\newcommand{\bu}{$\bullet$}
\newcommand{\ci}{$\circ$}

\newcommand{\hsi}{\hspace*{\sistretch}}

\begin{table}
\begin{tabularx}{\columnwidth}{|c|lp{\cs\columnwidth}p{\cs\columnwidth}p{\cs\columnwidth}p{\cs\columnwidth}|}
\hline
& \textbf{Requirements} & \textbf{R1} & \textbf{R2} & \textbf{R3} & \textbf{R4}  \\
& Sub-items & a\hspace*{\sistretch}b\hspace*{\sistretch}c & a\hspace*{\sistretch}b\hspace*{\sistretch}c    & a\hspace*{\sistretch}b\hspace*{\sistretch}c    & a\hspace*{\sistretch}b\hspace*{\sistretch}c\hspace*{\sistretch}d     \\
\hline
\parbox[t]{2mm}{\multirow{4}{*}{\rotatebox[origin=c]{90}{\textbf{Research\ \ \ }}}}
& ConceptCloud & \bu\hsi\ci\hsi\bu & \ci\hsi\ci\hsi\ci & \bu\hsi\ci\hsi\ci & \ci\hsi\ci\hsi\ci\bst\bu  \\ %for alignment (don't know why...)
& GitGraph     & \ci\hsi\ci\hsi\ci & \bu\hsi\ci\hsi\ci & \bu\hsi\ci\hsi\ci & \ci\hsi\ci\hsi\ci\hsi\ci  \\ 
& Linvis       & \bu\hsi\bu\hsi\ci & \ci\hsi\bu\hsi\ci & \bu\hsi\ci\hsi\bu & \ci\hsi\ci\hsi\ci\bst\ci \\
& RepoVis      & \ci\hsi\ci\hsi\ci & \ci\hsi\ci\hsi\ci & \bu\hsi\bu\hsi\bu & \ci\hsi\ci\hsi\ci\bst\bu \\
& Visgi        & \bu\hsi\ci\hsi\bu & \bu\hsi\bu\hsi\ci & \ci\hsi\ci\hsi\ci & \ci\hsi\bu\hsi\bu\bst\bu \\
\hline
\parbox[t]{2mm}{\multirow{4}{*}{\rotatebox[origin=c]{90}{\textbf{Git client\ \ \ }}}}
& GitHub               & \bu\hsi\ci\hsi\ci & \bu\hsi\bu\hsi\ci & \bu\hsi\bu\hsi\bu & \bu\hsi\bu\hsi\ci\bst\bu \\
& gmaster              & \ci\hsi\ci\hsi\ci & \bu\hsi\bu\hsi\bu & \bu\hsi\bu\hsi\bu & \bu\hsi\ci\hsi\ci\bst\ci \\
& gitk                 & \ci\hsi\ci\hsi\ci & \bu\hsi\bu\hsi\ci & \ci\hsi\bu\hsi\bu & \ci\hsi\ci\hsi\ci\bst\ci \\
& gitkraken            & \ci\hsi\ci\hsi\ci & \bu\hsi\bu\hsi\ci & \ci\hsi\bu\hsi\bu & \ci\hsi\ci\hsi\ci\bst\ci \\
& Sourcetree           & \ci\hsi\ci\hsi\ci & \bu\hsi\bu\hsi\ci & \bu\hsi\ci\hsi\bu & \ci\hsi\ci\hsi\ci\bst\ci \\

\hline
\end{tabularx}
\hspace*{1mm}
\label{table:assessment}
\caption{
    Assessment of related systems against the requirements. 
    %The number of black marks depends on the number of satisfied sub-item.
    The black mark implies that the related system satisfies the corresponding sub-item.
}
\end{table}

\subsection{Requirement Analysis}
Based on the final four tasks, we formulated the requirements that Githru must satisfy as follows: 
% \begin{itemize}[noitemsep, label={}, leftmargin=*]
% \setlist{nolistsep}
\setlist{topsep=0.1em}
\begin{itemize}[itemsep=1pt, leftmargin=1.5em,]
    \item \textbf{R1: Provide an overview.}
    The system should present an overview of development history where (a) the commits are grouped according to specific criteria to avoid examining each commit individually; (b) the visualization of a group encodes its size and topological position compared to others; and (c) the summary of the selected group(s) is presented interactively (T1).
    
    \item \textbf{R2: Visualize a graph while preserving topology.}
    The graph representing the abstracted data should be visualized in an interpretable form.  
    The graph should contain abstracted topological data that include (a) the temporal sequence of each node (i.e., commit) and (b) branch information and merge relation; and (c) the graph should be navigable with minimal interactions (T2).
    
    \item \textbf{R3: Support filtering by and searching for details.}
    Depending on the user query, which can be a keyword or a temporal range, the corresponding commits should be (a) filtered in or out and (b) searched and highlighted to reduce the exploration scope.
    Moreover, users should be able to (c) browse the details of each commit (T3).
    
    \item \textbf{R4: Support comparison.}
    The system should facilitate comparisons (a) based on the number of commits and LOC. The magnitude can be compared according to (b) overall trends, or (c) within/between user-selections.
    (d) In particular, the information in the changed files should be compared while being organized according to the directory that contains the structure of the source code (T3).
    
\end{itemize}

As discussed in Section 3, there have been various Git clients and research. 
Among them, we selected the ones that support visual exploratory analysis of the Git metadata for assessment.
We evaluated whether they meet the requirements (Table 1).
GitHub fulfills the requirements mostly due to \emph{GitHub Insights}, the tool with which in-field developers are most familiar.
Hence, we selected GitHub for comparison to Githru in the user study.

\section{The Githru System}

\begin{figure}[t]
    \centering
    \includegraphics[width=\linewidth]{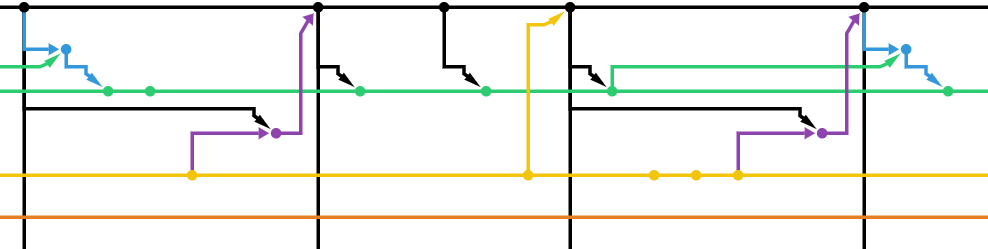}
    \caption{Complex git DAG (Captured from GitHub \cite{github} network graph of \emph{realm-java} \cite{realmJava} repository)}
    \label{fig:complex_dag}
\end{figure}

\begin{figure*}[t]
    \centering
    \includegraphics[width=\textwidth]{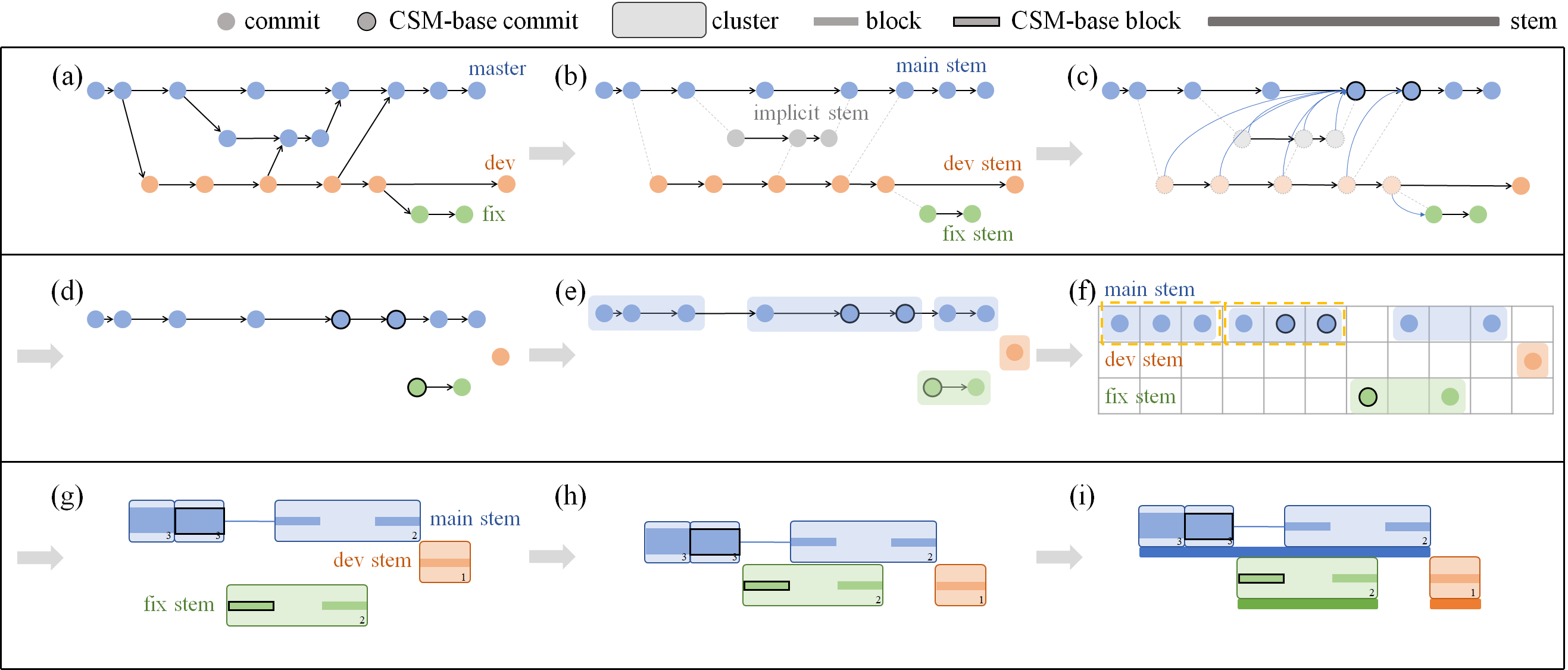}
    \caption{
        Git graph data abstraction from DAG to \emph{stem} structure.
        (a) DAG representation of a Git repository.
        (b) \emph{Stem} pruning simplifies edges, but produces implicit stems.
        (c) Performing a Context-Preserving Squash Merge (CSM).
        (d) The \emph{stem} structure reduced by CSM.
        (e) Commit clustering.
        (f) Group adjacent commits in the grid layout to the \emph{block}s (denoted by orange dashed box).
        (g) Squeezing the \emph{block}s.
        (h) \emph{Stem} relocation for space efficiency.
        (i) Adding a thin strip at the bottom of each \emph{stem}.
    }
    \label{fig:branch_2_stem}

\end{figure*}

We propose Githru, a Git repository visual analysis system, for users who are not acquainted with the underlying source code of a project.
This section describes data reconstruction techniques for resolving the visual complexity of the huge sets of Git metadata and novel visualization techniques for exploratory analysis of the metadata.

\subsection{Git Graph Data Abstraction}
The DAG representation of a Git repository suffers not only from a large number of nodes (i.e., commits) but also from diverging and converging links at implicit and explicit branches (Fig. \ref{fig:complex_dag}).
As the number of commits and branches inevitably increases over time in an ongoing project, scalability is crucial for DAG-based visual analysis.
As a remedy, we introduce graph reorganizing techniques tailored to the Git metadata, which could interactively reduce the number of nodes and links during analysis (R2).

\subsubsection{Transforming Branches to Stems}
The top straight line in the DAG of a Git repository generally represents the master branch (Fig. \ref{fig:branch_2_stem}a). However, an overwhelming number of branches and the connected links between them could hinder tracking down the origin of changes even for commits in the master branch.
To alleviate this problem, Githru removes the connected links between the branches in a DAG to form a group of \emph{stem}s (R1).
A \emph{stem} is a list of ancestor nodes for a specific commit that includes only one of the parents when there are multiple preceding nodes.
It is similar to the first-parent option of the \texttt{git log} command \cite{gitLogDocumentation}, which removes other parent nodes from a branch.
However, git log focuses only on a single branch while neglecting pruned nodes and their parents.
Conversely, Githru applies the approach to every branch to provide an overview of the overall history of development.

The process starts with building the main \emph{stem} from the \emph{master} branch, into which commits finally merge.
Pruning a branch could affect the topology of other branches that exclusively occupy common ancestor nodes.
Thus, we begin the process from the master branch to preserve the order of events in the mainline of development.
The rest of the branches are pruned afterward by retaining only the first parent commits in each branch.
Then, we remove links to non-first-parent nodes in adjacent \emph{stem}s to reduce visual complexity.
Eventually, only one path remains for every \emph{stem}.
Due to the simple topology and reduced number of edges, the result provides a brief overview of branches and enables simple traversal without any backtracking to multiple parents.

The downside of converting to a \emph{stem} structure is leaving extra implicit \emph{stem}s that have no branch information, as shown in Fig. \ref{fig:branch_2_stem}b.
Also, the process removes links between \emph{stem}s that hold branching and merging information.
However, in the case of understanding the context of development history, the experts in requirement analysis confirmed that they were interested in finding the contents of merged commits rather than the underlying links between branches.
We combat these disadvantages through a Context-Preserving Squash Merge, described in the following section.

\subsubsection{Context-Preserving Squash Merge (CSM)}
The preliminary interviews with domain experts indicated that tracking back from a merge commit to its parents is laborious and tedious in a DAG with many branches.
Transforming the DAG into a \emph{stem} structure could unravel the complexity with a straightforward topology, but it does not reduce the number of \emph{stem}s.
Moreover, it hides trails to the relevant \emph{stem}s and leaves a merge commit only for reference.
However, the content of a merge commit is insufficient if there is only a short sentence written by its committer or an auto-generated message \cite{michaud2016recovering}.

As a resolution, we propose a Context-preserving Squash Merge (CSM).
CSM fuses relevant commits (i.e., second parent commits from the merged branch \cite{10.5555/2695634}) into a single node for simplicity (Fig. \ref{fig:branch_2_stem}c) and fetches messages from the \emph{stem}s to preserve the merge context (R2).
For each merge commit on the main \emph{stem} (i.e., CSM-base), CSM traverses every parent commit (i.e., the CSM-source) on the other stems.
When a commit is a parent of multiple CSM-bases, we select the leftmost commit as a base to avoid redundant merges.
CSM gathers contextual information from every CSM-source (e.g., author, commit type, and log message) and appends it to the end of the corresponding field in the CSM-base.
For instance, the authors of CSM-sources become coauthors of the corresponding CSM-base.
However, the list of changed files remains the same since CSM-base encompasses the changes from CSM-sources.

In case of merged PRs, we also include additional information (e.g., pull request number, message, and body) from the PRs to comply with the comments from the preliminary interviews. 
After the CSM is applied to the main \emph{stem}, it is iteratively applied to the other \emph{stem}s, starting with the one with the most recent commit.
We determine the processing order to preserve the topology of critical stems first, following the valuation criteria in a prior work \cite{wessel2019should}.

The CSM can drastically reduce the number of \emph{stem}s and commits on the screen by removing implicit \emph{stem}s (Fig. \ref{fig:branch_2_stem}d), \emph{stem}s corresponding to merged branches, or merged PRs (R1). 
LinVis \cite{wilde2018merge} proposed a similar approach that grouped parents into a hierarchical structure and presented the structure in Merge-Tree.
However, this prior work focused mainly on analyzing the details of CSM-sources (i.e., parent commits) in the master branch and presenting a visual representation of the hierarchy.
In contrast, Githru provides an overview of the entire repository by applying the CSM to every \emph{stem}.
Furthermore, users can decide whether to apply a CSM or not, depending on their task.
Users can also, if necessary, explicitly visualize the edges between the CSM-base and CSM-sources.

\begin{figure}[b]
    \centering
    \includegraphics[width=\linewidth]{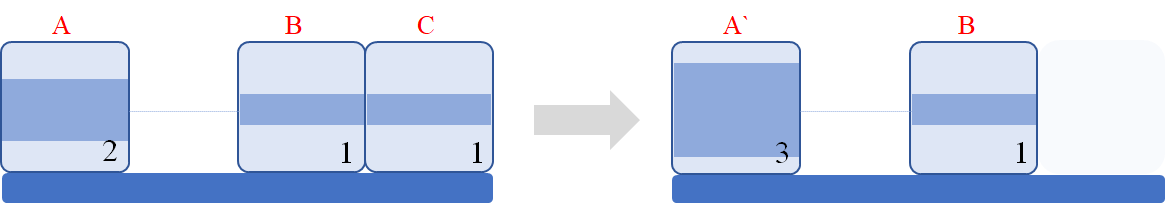}
    \caption{Clustering non-conflict commits simplifies the \emph{stem} graph by grouping similar non-neighbor commits.}
    \label{fig:nonconflict-clustering}
\end{figure}

\subsubsection{Commit Clustering}
In another effort to improve scalability, we adopt a clustering technique to group neighboring commits in each \emph{stem} (Fig. \ref{fig:branch_2_stem}e; R1).
The scope of the grouping is confined to similar commits in each \emph{stem} to preserve the temporal sequence and topology.
We exploit the Simple Additive Weighting (SAW) model in calculating similarity since this model is intuitive for users to understand and is known to serve exploration well \cite{wall2017podium, zanakis1998multi}. 
To measure similarity, we choose five criteria from the commit metadata (i.e., author, commit date, commit type, file, and message).
Jaccard similarity is used for the author, commit type, and file. Cosine similarity is used with TF-IDF weights for the message, and logarithmic growth from 0 to 1 is used for the commit date.
Users can interactively change the granularity clusters by adjusting the similarity threshold. %; if the similarity score is higher than the threshold, then clustered.
We also prepare options to determine whether to separate commits from different release versions \cite{semanticVersioning2}.
This enables separate analyses of individual releases, which was found to be significant in the requirement analysis (R3).

If there is still an overwhelming number of nodes even after the above techniques are applied, users can additionally apply Non-Conflict Commits Clustering, which can group non-neighbor commits (R1).
For instance, suppose that a cluster A is not adjacent to a cluster C, but their similarity is above the threshold (being sufficiently similar). 
If the cluster A has no commonly modified files with cluster B (an in-between cluster of A and C), changing the order of B and C could further simplify the underlying structure by grouping the clusters A and C as illustrated in Fig. \ref{fig:nonconflict-clustering}.
However, we make this process optional because it considers only the conflict coming from modified files and not the contextual conflict.

\begin{figure}[t]
    \centering
    \includegraphics[width=\linewidth]{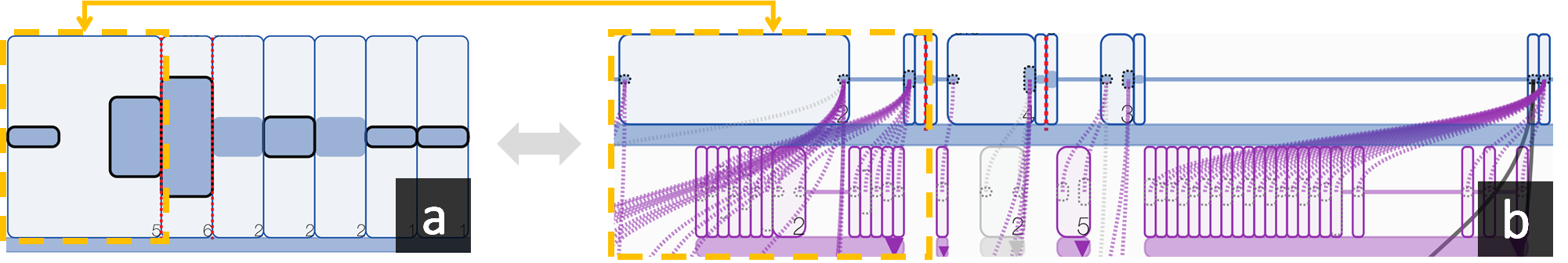}
    \caption{
    Users can (a) apply the CSM or (b) not. 
    Without the CSM, the clusters containing CSM-bases will be split due to the \emph{block}s of newly appeared CSM-sources, as depicted in (b).}
    \label{fig:csm_disable}
\end{figure}

\subsection{Stem Graph Visualization}
Representing a cluster (i.e., commits grouped by similarity) as a single node saves screen space by reducing the number of nodes.
However, it might also destroy the temporal order between the commits that belong to different clusters in different stems. %as depicted in Figure \ref{fig:cluster_and_block} (a).
Thus, we introduce a \emph{block} in the cluster visualization as follows:
% to complement both reducing search space and preserving temporal sequence of clusters.

% \setlist{nolistsep}
\setlist{topsep=0.1em}
\begin{enumerate}[noitemsep, leftmargin=*]
\item A two-dimensional grid layout is drawn in which columns and rows are mapped to time slots and \emph{stem}s, respectively.
\item Each commit is filled in a cell in temporal order. 
\item Adjacent commits are grouped into a \emph{block} until the \emph{block} overlaps with any commit in other \emph{stem}s.
\item The \emph{block} is squeezed into a single column (Fig. \ref{fig:branch_2_stem}f).
\item A vertically centered box is drawn for each \emph{block}, where the height of the box corresponds to the number of commits in the \emph{block}.
\item An outline is drawn to enclose the \emph{blocks} in a cluster, along with the number of commits included in the bottom-right corner (Fig. \ref{fig:branch_2_stem}g) when the rendered cluster has sufficient width.
\end{enumerate}

We emphasize each \emph{block} using a black border line when it has a CSM-base and the CSM is enabled (Fig. \ref{fig:branch_2_stem}g); the line turns dashed gray if the CSM is disabled (Fig. \ref{fig:csm_disable}).
Regarding visual clutter of borders, adjacent clusters with identical pale colors were hardly distinguishable without any additional visual cues. This issue was raised during the interviews with domain experts, and we eventually included borders.
The visibility of the edges between the CSM-base and CSM-sources also changes accordingly.
This allows us to reduce the number of visual elements in the horizontal dimension without losing the temporal order of commits across \emph{stem}s (R1, R2).

We also work on optimizing the vertical space by filling the gaps in the grid.
In the aforementioned grid layout, each row maps to each \emph{stem}.
The main \emph{stem} comes first, on the top, and the other \emph{stem}s follow according to the date of the last commit in descending order.
While such a layout is intuitive, the height of the grid increases as the number of \emph{stem}s grows.
To alleviate this problem, we relocate \emph{stem}s that do not overlap with each other into a single row (Fig. \ref{fig:branch_2_stem}h).

However, such relocation can cause difficulty distinguishing between individual \emph{stem}s in the same row.
Thus, we introduce various visual aids.
We add a thin strip under each \emph{stem} that not only helps differentiate the \emph{stem}s but also allows entire clusters in the \emph{stem} to be selected at once (Fig. \ref{fig:branch_2_stem}i).
We then move the \emph{block}s inside each \emph{stem} to the vertical center to avoid interference with short \emph{blocks}.
We also add an edge between non-adjacent clusters in a \emph{stem} to indicate they belong to the same \emph{stem}.
And if a \emph{block} includes a commit with a release tag, we draw a red dashed line on the right of the block with the version number on top.
The final \emph{stem} abstraction is shown in Fig. \ref{fig:teaser}c.

\subsection{Cluster Visualization}

Following Shneiderman's mantra \cite{shneiderman1996eyes}, we prepare two levels of detail for cluster views.
The views are designed to support commit-level analytical tasks, which were demanded in the interview.
Users can start by selecting one or more clusters in the \emph{stem} graph to focus on specific details.
However, the selected clusters can have a varying number of commits (ranging from a single commit to all commits in the repository), and each commit contains intricate information (e.g., author, keywords, and list of modified files) to visualize in a single view.
Thus, we present underlying details in a stepwise manner with two coordinated views.

\subsubsection{Grouped Summary View}
Grouped Summary View shows a brief overview of the selected clusters as shown in Fig. \ref{fig:teaser}d (R1).
The columns in the view are mapped to individual clusters and the width of each column is proportional to the number of commits.
This view enables a visual comparison of the relative size among selected clusters, which was frequently cited as a needed task in the requirement analysis (R4).
Each column has a group of horizontal bars that briefly show the top two or three values from the clustering criteria (i.e., author, commit types, modified files, and keywords).
In addition, there are bars for the list of modified directories and files to offer more context.
Furthermore, the length of each bar is proportional to the number of relevant commits that users could visually compare.
For instance, users could find the author who has contributed the most to the cluster by finding the longest bar.

Enabling the Summary by CLOC option changes the width of each column and the length of the file criteria bar proportionally to the number of CLOCs (changed LOCs, added LOCs + deleted LOCs).

\subsubsection{Cluster Detail View}
When users select a cluster in Grouped Summary View, Cluster Detail View appears at the bottom.
This view provides commit-level details along with a visual summary of the affected files and directories on the left (R1).
A list of raw commit metadata is presented in a tabular form by date in ascending order (Fig. \ref{fig:teaser}f).
In the case of a CSM commit, it shows only the CSM-base at first, but users can expand the row to also see the relevant CSM-source commits.
On the left of the table, we prepared a file icicle tree \cite{heer2010tour} (Fig. \ref{fig:teaser}e).
Since files and directories are organized in a hierarchy, we consider a number of space-filling approaches to maximize space utilization \cite{johnson1991tree}.
Among Tree-Map \cite{johnson1991tree}, SunBurst \cite{stasko2000evaluation}, and the icicle tree, we finally choose the last to comply with the task requirements.
Tree-Map shows limitations in the structural interpretation task \cite{bladh2004extending} and SunBurst is inadequate to embed long file names because of the radial coordinate.
On the other hand, the icicle tree explicitly shows a structural hierarchy in a Cartesian coordinate system well suited for displaying a string (e.g., file-name) horizontally \cite{heer2010tour}.
Also, as the depth of the modified file structure can vary, we enable users to zoom in or out with a mouse click on the icicle tree (R3).

\subsection{Controlling the Analysis Scope}

We provide additional visual components to facilitate the in-depth analysis by controlling its scope: Cluster Parameters, Filter, and Search.

\subsubsection{Cluster Parameters}
One can control the granularity of clustering by setting a Clustering Step (Fig. \ref{fig:teaser}b) encoded as a vertical slider.
The desired level of abstraction can be set by adjusting the maximum difference value (threshold) to be clustered. For instance, if one moves up the slider, the clustering becomes granular.
Thus, one can analyze fine-grained clusters by moving up the slider.
For the same reason, we also provided a way to set Preference Weights for each similarity criterion. For instance, if one wants to cluster only commits with similar commit types, one can simply set the weight of the commit type to 1 and the rest to 0. Such a capability reveals the underlying policy of clustering, helping users to understand the context (R2) and find the information they want (R3) by allowing them to set appropriate clustering schemes for their task.

\subsubsection{Filter and Search}

\vspace{-0.05cm}
\parahead{ \ \ \ \ Global Temporal Filter} 
In many tasks, understanding what happened over a period of time is important. 
For example, a manager may wonder what happened over the previous week or month. Therefore, we provided a Global Temporal Filter with ways to filter for a certain time period: Brushing (Fig. \ref{fig:teaser}a) and a Select Box.
Githru provides two horizontal bars that can be brushed. 
The bar at the top includes two area charts aligned vertically, which represent the number of commits and LOCs by date respectively. 
The bar at the bottom is a horizontal list of boxes that encodes each commit ordered by date. 
Both brushes allow for filtering in a specific range and they are synchronized.
However, the interval of the commits is non-uniform, unlike the dates in the above chart, so we draw a line connecting the release commit to the bin, pointing to the release date as a guideline. This can alleviate the confusion that occurs when the range of the top brushed area differs from that of the bottom brushed area. 
This method allows users to effectively select a specific period of dates or commits (R3).
Users can also select a specific date or release tag using the Select Box. This solves the problem that occurs in brushing when the user has to choose an exact position, which is difficult to select.

\parahead{Keyword Filter} 
Users often have to work on tasks related to certain elements. For example, if a newcomer replaces Alice in a project, the newcomer may want to see only what Alice did. For such cases, Githru offers the Keyword Filter, which can exclude or include the commits related to specific keywords.
We provide keyword-filtering for all similarity criteria except for commit date, which is filtered by the Global Temporal Filter.

\parahead{Stem Type Filter}
Each \emph{stem} type in Githru has various characteristics, such as the existence of a name, its relation to PRs, and its PR status. Users may want to focus on a particular \emph{stem} type depending on their task. For instance, there is no need to see merged or closed branches when looking for recently opened PRs. Thus, we offered options to show or hide each \emph{stem} type (R3).

\parahead{Search and Highlight}
If one searches for a certain keyword, Githru scans branch names, tags, commit messages, authors, commit IDs, and modified files. Then, it highlights every \emph{block} that matches. Multiple keyword highlighting is also allowed  (R3).

\subsection{Comparison View}
Comparison View provides a detailed comparison of clusters or cluster sets (R4) based on similarity criteria and \emph{stem} topology. It is designed following the details-on-detail strategy: a rough comparison in Grouped Summary View, and a detailed comparison in Comparison View. We used keywords instead of raw messages for comparison since it was more difficult to use unstructured strings than keywords when visualizing the differences and commonalities between clusters. We used Selection Cards to compare \emph{stem} information and date and used a Diff View for the other criteria.

% \begin{figure}[t]
%     \centering
%     \includegraphics[width=\linewidth]{vgtc_journal_latex/figures/compare_tagged.png}
%     \caption{Compare View provides the comparison by a) Authors \& Commit type, b) Files, and c) Keywords.}
%     \label{fig:compare}
% \end{figure}

\parahead{Capture Selection}
Users can select the clusters in Grouped Summary View as a selection to be compared. To enhance reusability, we provided selection capturing to select once and use it repeatedly.

\parahead{Selection Cards}
Selection Cards represent corresponding selections. 
The \emph{stem} information, such as a branch name or PR number, is prominently presented on each card, as they directly represent the characteristics of the cluster (R1). For the same reason, the color of the card is also derived from the color of the \emph{stem} where the selected cluster(s) is located.

\parahead{Diff View}
Diff View shows a two-way comparison between selections for authors, commit types, files, and keywords (Fig. \ref{fig:teaser}g). Since comparison becomes difficult as the number of objects increases \cite{gleicher2017considerations} and Grouped Summary View already provides a rough overview of a multi-way comparison, a two-way comparison fits the details-on-demand strategy (R3, R4). 

Diff View assigns different colors to three sets: the intersection, the difference of A from B, and the difference of B from A to perform a two-way comparison more effectively. Moreover, the view provides the option to hide or show elements in each set so that users can focus on a particular set(s). For instance, if one wants to see only the intersection between two clusters, one needs to deselect only the parts representing the difference.

As for the author and commit type criteria, we displayed all values: the number of developers for a repository is typically below ten due to the internal guidance gathered during prior interviews. Hence, the top ten elements for others are represented.
Furthermore, to ensure consistency with Grouped Summary View and to offer more information to users, we provide an option to select between number of commits (commit \#) and CLOC for comparison. Diff View is composed of three parts as follows:

\setlist{topsep=0.1em}
\begin{itemize}[noitemsep, leftmargin=2em]
% \begin{itemize}[label={}, leftmargin=*]
% \begin{itemize}[noitemsep, label={}, leftmargin=*]
% \begin{itemize}
    \item Authors \& Commit type: Display a stacked bar chart representing each cluster. Each box represents the value (CLOC or commit \#) of the corresponding author.
    \vspace{0.05cm}
    \item Files: Present individual bar charts for two clusters while showing the top ten modified (CLOC or commit \#) files in each cluster. The bar size encodes \{value / total value (of cluster)\} where value is CLOC or commit \#, depending on the selected option.
    \vspace{0.05cm}
    \item Keywords: Visualize the top 20 words ranked by TF-IDF value as a word cloud. As the statistical overview of the word cloud is achieved by positively correlating the font size \cite{heimerl2014word}, we set the size of the words in the word cloud at a minmax normalized TF-IDF value. Thus, the user can perceive the importance of each word.
\end{itemize}

\subsection{System Architecture and Implementation}
The Githru system consists of two modules - pre-processing and visualization. 
The former module crawls and extracts metadata (i.e., list of commits and PRs) from the Git repository using the GitHub API in Python.
It extracts keywords from each commit message and removes noises (e.g., stop words, enumeration symbols) using the Natural Language Toolkit Library \cite{bird2009natural}.
It also calculates the TF-IDF weight of keywords and the similarity between commits in advance for faster performance at runtime.
The latter module shows an interactive visualization of the extracted data as a single-page application in JavaScript using D3.js and the React library.
The left pane consists of the \emph{stem} graph, detailed views and related control panels, and the right pane shows the Comparison View.
It is best suited for full-screen mode in FHD (1920X1080) resolution.
\section{Evaluation}

We consulted the infrastructure team at Samsung to find well-managed Git repositories, which staff maintained actively for a commercial product.
They recommended the \emph{A-project} in-house repository, for which team members had committed actively, used release tags properly, and wrote well-organized messages.
As another dataset, we selected the public GitHub repository \emph{realm-java}, which satisfies the same criteria.
We conducted a qualitative study as well as a quantitative study with the datasets.

\subsection{Qualitative Study}
To assess the efficacy and usability of Githru, we performed case studies with the \emph{A-project} and the \emph{realm-java} GitHub repository, followed by interviews with domain experts.
This study was conducted at Samsung Research with four previous interviewees (D3, L1--3) and one manager (M1) for about an hour and a half each. They had an average of 11.4 years (ranging from 7 to 16 years) of professional software development experience.
We demonstrated the Githru system and discussed its effectiveness and usability.

\begin{figure}[t]
    \centering
    \includegraphics[width=\linewidth]{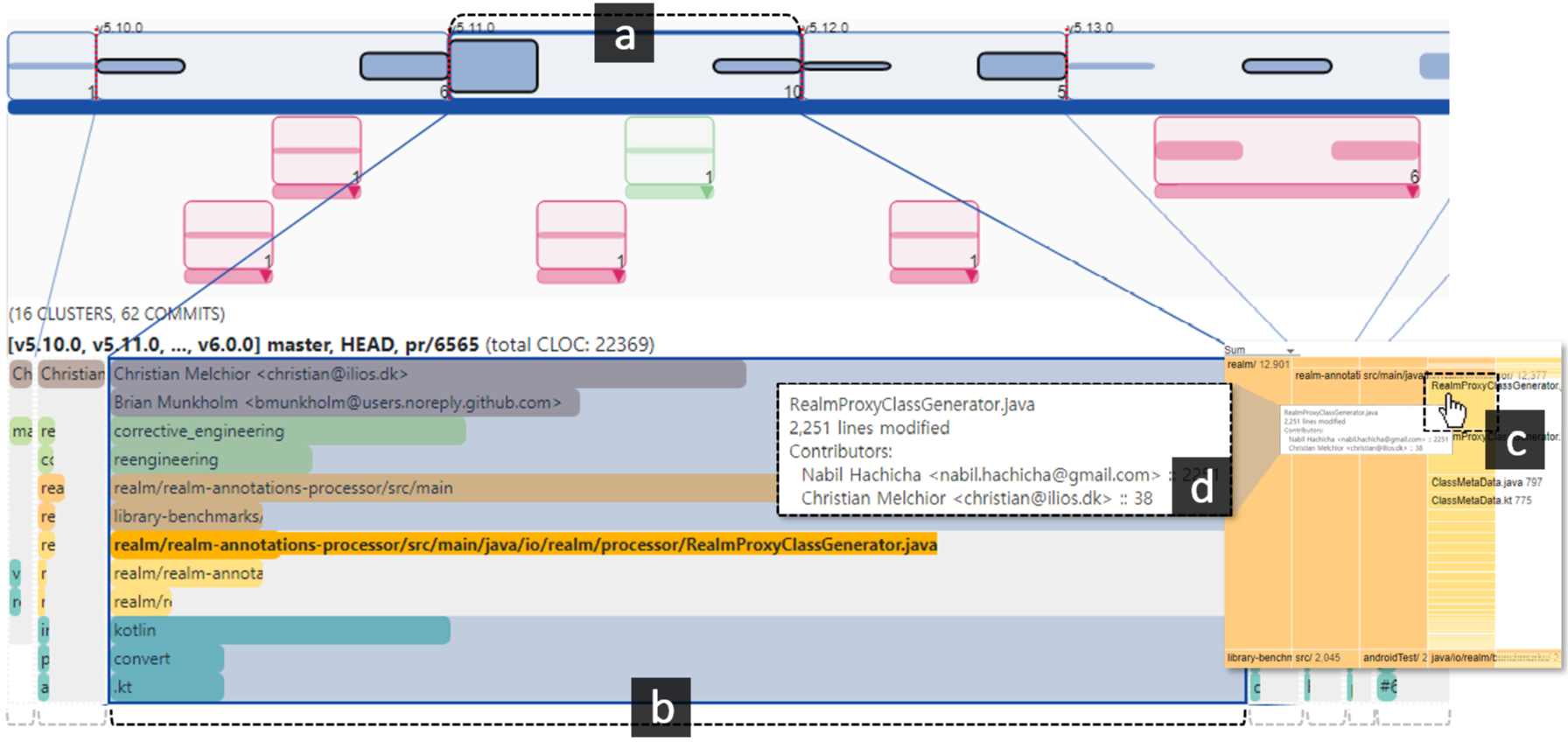}
    \caption{
        Users can easily identify a hotspot release and the file with the highest CLOC contributions.
        (a) Cluster for release 5.12.0.
        (b) Cluster with the highest CLOC.
        (c) File with the highest CLOC.
        (d) Tooltip for file information.
    }
    \label{fig:case_study}
\end{figure}

\parahead{Exploration and Overview}
All of the experts started to use Githru by exploring the \emph{stem}s or clusters and by looking into their summaries in Grouped Summary View.
Then they used the Global Temporal Filter to focus on different time periods.
They also interactively adjusted the clustering parameters to find suitable granularity, which barely changed once fixed.
All of them were especially interested in summaries about the authors.
They were thrilled to find who contributed the most by navigating through each cluster or \emph{stem}.
Then they repeatedly hovered the mouse cursor on some of the authors to check the trend of activities.
When an unexpected name appeared, they checked the list of commits in Cluster Detail View and explored the file icicle tree to see which modules the author worked on (Fig. \ref{fig:case_study}d).
D3 and L3 were satisfied with the commit type classification, although we admit that the accuracy of the classification still has room for improvement.
In some cases, they were surprised to find  \emph{forward} (feature) type commits on a certain release since the release was supposed to have fixes only.
They went through the commit list in the Cluster Detail View and could find the reason easily. If a big cluster appeared, they tried to split it up by adjusting clustering parameters (i.e., moving the Clustering Step slider to a smaller value). They changed Preference Weights to focus on fewer similarity criteria, which also led to finer granularity of clusters.

In terms of aiding exploration, L2 admired a displayed graph with PR information.
Actually, it had been requested by a domain expert during the preliminary interviews, as it was difficult to obtain an overview of PR information.
While GitHub provides the \emph{Pulse} page for this purpose, it has limited capability (i.e., only specific periods can be selected), which was the same as others in \emph{GitHub Insights}.
As a result, one has to visit each PR page one by one in GitHub.
To the best of our knowledge, Githru is the first tool to integrate PR information in an interactive overview along with relevant metadata.

L3 noted a real-world case in which Githru might come in handy.
When the company hires another company to work on a project, the project manager needs to monitor whether it is proceeding as planned.
Formerly, the manager from the hired company reported periodically about the progress of individual developers.
Thus, the report was not made in real-time and the management had to be passive.
However, L3 complimented that Githru can help the manager track the development history interactively, even on a more detailed scale.
In short, we were able to confirm the efficacy of Githru in real-world exploration tasks.

\parahead{Release}
Participants in the case study also complimented the exploration by \emph{release}.
Since \emph{release} serves as a milestone in the development cycle, the experts performed analysis on a release-by-release basis (Fig. \ref{fig:case_study}a).
They became especially enthusiastic about being able to identify commits or PRs that were merged into a specific release.
Previously, to retrieve the commits, they had to navigate the commit graph in git log, Sourcetree, or GitHub, but it was not easy to understand the topology with these tools.
The graphs not only suffer from scalability issues but also from limited interactions to traverse a huge and complex network.
However, in Githru, one can locate a specific release and select commits belonging to the release by simply brushing the Global Temporal Filter.
Then, one can disable CSM operation and hover over the specific cluster to see corresponding commits.

L1--3 and M1 stated that this capability would be useful for writing release notes.
They said existing auto-generation techniques for release notes were seldom used due to their incompleteness, so developers had to write them manually.
The process demanded a laborious effort to traverse all commits reflected in the release and sometimes caused errors such as missing certain information.
We actually discovered such an error in the in-house dataset using Githru; a specific release note mistakenly included information from commits that were already reflected in the previous release.
Some experts suggested that it would be helpful in writing a release note draft or a weekly report if one could customize the format of the commit list.

\parahead{File Information}
During prior interviews, managers were mainly interested in finding a hotspot.
This aligns with the finding in prior work that a hotspot could indicate defects and complexity of code \cite{hassan2005top, kim2007predicting, nagappan2005use}.
The managers also mentioned that a tester could handle test cases related to the hotspot as suggested in prior work \cite{tao2012software}.
The experts started looking for a hotspot by comparing the width of the summary boxes.
Then, they selected the cluster corresponding to the widest column so they could narrow down the search space (Fig. \ref{fig:case_study}b). Finally, they went through the file icicle tree in the detailed summary (Fig. \ref{fig:case_study}c).
Unexpectedly, one of the experts found that the most changed files in terms of CLOC in the project were mainly big json data files.
He also found the person worked on the file by simply hovering on the node of the icicle tree.
All the experts confirmed that the icicle tree was an informative and intuitive way to analyze the hotspot, which is defined as frequently changed files.

\parahead{Understanding Stem Topology}
In the \emph{stem} graph, the experts found stale \emph{stem}s without any recent commits that had not merged into the main stem.
Unlike the other participants, L3 wanted to manage and track the stale \emph{stem}.
Thus, L3 tried to figure out why the stale \emph{stem}s remained and found several cases: freezing a specific version, maintaining documents (e.g., GitHub pages), and just left unmanaged for various reasons.
L3 was inclined to use \emph{stem}-based visualization since it was closely coordinated with other views, which served the context.
L2, L3, and M1 were interested in the state of PR \emph{stem}s. 
They used a similar approach to what was used to find stale \emph{stem}s to explore the PR \emph{stem}s and find out why an old PR was still open. 

In the prior interview, L3 had specifically requested to be able to try Githru for an upcoming \emph{B-project}.
However, the \emph{B-project} was known to have some notorious problems: there had been no manager, few documents for reference, and little collaboration between members.
L3 had tried to get an overview of its history with existing tools, but experienced difficulty exploring the context due to its overwhelming complexity.
Thus, we demonstrated examination of the \emph{B-project}'s repository to L3 during the case study.
It took only a couple of minutes to come up with the following insights: Each member was using their own branch without merging it into the master, they were working on independent modules, and a member who recently joined the project had started to merge all existing branches into the master using pull requests.
L3 was able to identify the complex structure at a glance, but could not find any interesting underlying details due to the overwhelmingly complicated structure.
Still, L3 was satisfied with the experience and reported that the insights were helpful for preparation.

\parahead{Filtering and Searching}
D3 and L3 greatly appreciated the filter and search features.
They started using the features without hesitation and reported them as two of the most valuable features in Githru.
According to our survey results, most Git clients had a search or filter function, but these show limitations when compared to Githru.
Existing tools implemented only one of the two features. 
Moreover, there is a trade-off between the two features: 
If the filter is applied, non-matched commits are removed from the graph, and the result is shown in a compact way. However, the list loses the graph representation that includes branch information (i.e., Sourcetree, gmaster, GitHub). Conversely, searching normally emphasizes matched commits or blurs non-matched commits, so it retains the graph representation. 
However, the number of displayed commits is not reduced, so users must keep manipulating the next match button and cannot grasp the results at a glance (i.e., gitk, gitkraken). 

Githru overcame the trade-off between filtering and searching.
Not only did we implement the two features, but we also ensured that filtering in Githru does not result in losing the graph information. Moreover, clustering by similarity is still possible thanks to the simplicity of the graph from the \emph{stem} structure. Therefore, the experts could do the same tasks regardless of filter and search.

\subsection{Quantitative Study}

To confirm Githru's effectiveness and usability, we recruited 12 developers (ten men and two women) with 3 to 15 years (average: 7.75) of development experience at Samsung and conducted a controlled user study. As we used an \emph{A-project} dataset, we excluded any participant who had experience with the project. The participants were from various teams working on cloud, UX, AI, process, device driver, and other projects. 

We designed three questions to ensure that Githru satisfies the requirements. Detailed descriptions of each question and their corresponding requirements are as follows:

\setlist{topsep=0.1em}
\begin{itemize}[itemsep=1pt, leftmargin=1.5em]
    \item Hotspot: Which file has the largest CLOC (most changed) over a specific period? (R1(a, b, c), R2(a, c), R3(a, c), R4(a, b, c, d))
    \item Comparison: Name the differences and commonalities of the author set who worked on the commits included in the two periods (R1(a), R2(a, c), R3(a, c), R4(a, c)).
    \item Topology: What is the first release that reflects a specific PR? (R1(a), R2(a, b, c), R3(a, b, c)).
\end{itemize}

As mentioned in Section 4, we selected GitHub for the comparison with Githru. We also let participants use git log with GitHub, since it is the most basic tool for developers and provides the richest raw data. Furthermore, in the prior interview, many developers said they view repository information in the git log. Therefore, we designed the experiment to compare Githru against GitHub and git log.

\begin{figure}[t]
    \centering
    \includegraphics[width=\linewidth]{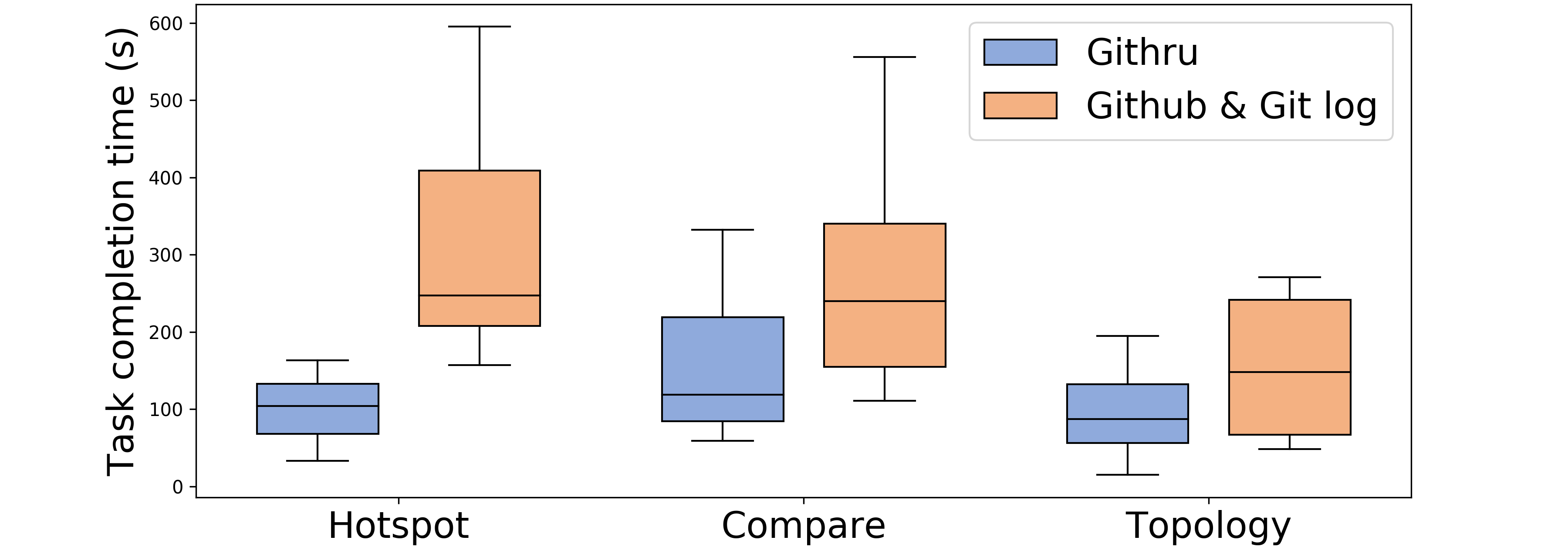}
    \caption{Quantitative evaluation result; the task completion time spent by subjects on three questions }
    \label{fig:user_study}
\end{figure}

The experiment was conducted as a within-subject study. The total experiment time was 50 to 60 minutes, which included 20 minutes of prior explanation and 5 minutes of tool demonstration and practice.
We created two similar problems for each question and the subjects solved them, in turn using both Githru and GitHub/git log. For git log, we guided the usage of the command parameters to solve the problems.
At this time, to prevent the order effect, half of the subjects used Githru first and the rest used GitHub/git log first.
We measured the time it took for subjects to give a correct answer by notifying them when they gave a wrong answer. The results show that Githru is more effective for solving problems related to hotspot $(Z = -5.76$, $p < .001)$, comparison $(Z = -3.78$, $p < .01)$, and topology $(Z = -2.69$, $p < .05)$ than GitHub/git log. 
The results are shown in Fig. \ref{fig:user_study}.

\subsection{Fulfillment of Requirements}
We also asked subjects to check several statements about whether the requirements were satisfied.
The survey was conducted using a 5-point Likert scale. The results show that the subjects agreed that Githru fulfills the requirements (average: $[$R1: 4.25, R2: 4.33, R3: 4.33, R4: 4.83$]$) and that the differences between the requirements are small enough. The subjects also agreed with the statement about usability (average: 4.17).

\section{Discussion}

\parahead{ \ \ \ \ Usefulness}
At the most basic level, all the interviewees appreciated Githru simply because it enables them to obtain an overview; existing tools have difficulty supporting even a simple aggregation of desired data.
In addition, we discussed another usage of Githru: if the development team's adoption of it in the real field.
L1 said that our system is also appropriate for exploring a project's overall history even when a user does not have concrete questions, such as Where should I start to explore from? or What do I want to find (author, file)? This means Githru is also effective for exploratory data analysis (EDA) of Git history.
So, if a new frequent pattern can be discovered during the EDA using Githru, the team can display it on the team dashboard to share the insight.

Also, Githru collaborates with existing tools. During analysis, it is helpful if existing tools are supported when detailed information is needed. For instance, the branch explorer of gmaster can reveal where the commit is branched or what the relationship is between non-main line branches. As for the details of a PR, GitHub is the only tool for reading conversations attached to a PR.

\parahead{Limitation and Improvements}
The stem structure forces every operation to execute horizontally except for the CSM.
As a result, we could work toward greater simplicity and better interpretation of a DAG. However, within the stem structure, we cannot perform vertical operations such as clustering two related stems and summarizing multiple stems at once.

We employed commit classification and topic analysis methods in this study. Our focus was not on the methods themselves, so lightweight baseline methods are selected as a proof-of-concept. High performance techniques, such as summarizing the messages of a cluster as short sentences or more precise classifications, may offer more efficacy to users.

During the study, we found some cases in which clustering by release did not work properly. This occurs when a release-tagged commit is not on the master branch, but only on a branch that was created solely for the release. In such a case, the commit does not merge with the master branch and the main stem will not hold the information even when the CSM is applied.
Githru supports changing the name of the main branch by URL parameter when the main branch is not the master. However, if the release branch continuously changes its name, the stem reorganization does not work properly. 

\parahead{Future Work}
The extension of metadata enables extensive and in-depth analysis as well.
Collecting datasets from heterogeneous systems such as an issue tracking system, CI (continuous integration), and test environments, and arranging them along the Git history will give us a holistic understanding of the overall development history.
Also, adding more information about source codes to metadata facilitates in-depth analysis. 
Currently, file information such as name, path, and CLOC is the closest to the source code among the Git metadata used in our study. 
Hence, adding extracted fine-grained source code change to the metadata will give detailed information about source codes (i.e., function, class, and edit operations) \cite{fluri2007change}.
In obtaining them, the synchronizations between heterogeneous data, scalability, and performance issues are the main obstacles to be solved.

During the case studies, the experts provided practical suggestions for supporting multiple repositories.
They are running multiple repositories for one project according to modules. For instance, an end-user device can have three repositories for device, smartphone, and server.
In that case, several issues occur when visualizing the metadata of three repositories at once such as three main lines, a bigger graph representation, and synchronization issues.
A huge project, for example, an OS platform such as Tizen \cite{tizen}, runs over a thousand Git repositories, so there can be serious scalability issues when visualizing the overall development history.
Also, the current Githru visualization module is running as a standalone front-end application, as mentioned before. If the number of commits greatly increases for the aforementioned reasons, it will be necessary to translate them to the server-client model.
Moreover, streaming commits are not supported due to the preprocessing step in Githru. However, we would be able to handle such data by revising the step to process incrementally. We leave these issues for future works.
\section{Conclusion}
We presented Githru as an interactive visual analytics system for the Git metadata to help users explore and understand the context of development history. 
The contribution of this paper is fourfold. 
First, we refined analytics tasks and system requirements for Git metadata through literature reviews and expert interviews. 
Second, we proposed novel techniques tailored to the metadata to abstract a large-scale Git commit graph: reorganization of a DAG to a stem structure, context-preserving squash merge methods, and clustering reduce the number of stems and commits. 
Third, Githru provides a summary view for clusters to grasp the overview and a comparison view with which users can compare different clusters. 
Lastly, we evaluated the system with real-world datasets and domain experts from a major international IT company.
Case studies with domain experts and a controlled user study with developers, comparing Githru to a combination of GitHub/git log, were conducted and confirmed the effectiveness and usability of Githru.
The implementation of our system is available at github.com/githru/githru.

\acknowledgments{

This work was supported by the National Research Foundation of Korea(NRF) grant funded by the Korea government(MSIT) (No. NRF-2019R1A2C2089062, No. NRF-2019R1A2C1088900) and by Korea Electric Power Corporation. (Grant number:R18XA01).
Young-Ho Kim was supported by Basic Science Research Program through the NRF funded by the Ministry of Education (NRF-2019R1A6A3A12031352).
The ICT at Seoul National
University provided research facilities for this study.
Hyunjoo song and Jinwook Seo are the corresponding authors.
The authors thank the participants from Samsung Electronics for interviews and evaluation.
}

\bibliographystyle{abbrv-doi}

\bibliography{githru}

\begin{thebibliography}{10}

\bibitem{gitLogDocumentation}
Git log documentation.
\newblock \url{https://git-scm.com/docs/git-log}.

\bibitem{github}
Github.
\newblock \url{https://github.com/}.

\bibitem{gitk}
gitk.
\newblock \url{https://www.atlassian.com/git/tutorials/gitk}.

\bibitem{gitkraken}
Gitkraken.
\newblock \url{https://www.gitkraken.com/}.

\bibitem{gmaster}
gmaster.
\newblock \url{https://gmaster.io/}.

\bibitem{realmJava}
realm-java.
\newblock \url{https://github.com/realm/realm-java}.

\bibitem{semanticVersioning2}
Semantic versioning 2.0.0.
\newblock \url{https://semver.org/}.

\bibitem{sourcetree}
Sourcetree.
\newblock \url{https://www.sourcetreeapp.com/}.

\bibitem{tizen}
tizen git.
\newblock \url{https://review.tizen.org/git/}.

\bibitem{alcocer2019performance}
J.~P.~S. Alcocer, F.~Beck, and A.~Bergel.
\newblock Performance evolution matrix: Visualizing performance variations
  along software versions.
\newblock In {\em 2019 Working Conference on Software Visualization (VISSOFT)},
  pp. 1--11. IEEE, 2019.

\bibitem{alexandru2019evo}
C.~V. Alexandru, S.~Proksch, P.~Behnamghader, and H.~C. Gall.
\newblock Evo-clocks: Software evolution at a glance.
\newblock In {\em 2019 Working Conference on Software Visualization (VISSOFT)},
  pp. 12--22. IEEE, 2019.

\bibitem{anslow2013sourcevis}
C.~Anslow, S.~Marshall, J.~Noble, and R.~Biddle.
\newblock Sourcevis: Collaborative software visualization for co-located
  environments.
\newblock In {\em 2013 First IEEE Working Conference on Software Visualization
  (VISSOFT)}, pp. 1--10. IEEE, 2013.

\bibitem{barik2015commit}
T.~Barik, K.~Lubick, and E.~Murphy-Hill.
\newblock Commit bubbles.
\newblock In {\em 2015 IEEE/ACM 37th IEEE International Conference on Software
  Engineering}, vol.~2, pp. 631--634. IEEE, 2015.

\bibitem{begel2014analyze}
A.~Begel and T.~Zimmermann.
\newblock Analyze this! 145 questions for data scientists in software
  engineering.
\newblock In {\em Proceedings of the 36th International Conference on Software
  Engineering}, pp. 12--23, 2014.

\bibitem{biazzini2014analyzing}
M.~Biazzini, M.~Monperrus, and B.~Baudry.
\newblock On analyzing the topology of commit histories in decentralized
  version control systems.
\newblock In {\em 2014 IEEE International Conference on Software Maintenance
  and Evolution}, pp. 261--270. IEEE, 2014.

\bibitem{bird2009promises}
C.~Bird, P.~C. Rigby, E.~T. Barr, D.~J. Hamilton, D.~M. German, and P.~Devanbu.
\newblock The promises and perils of mining git.
\newblock In {\em 2009 6th IEEE International Working Conference on Mining
  Software Repositories}, pp. 1--10. IEEE, 2009.

\bibitem{bird2009natural}
S.~Bird, E.~Klein, and E.~Loper.
\newblock {\em Natural language processing with Python: analyzing text with the
  natural language toolkit}.
\newblock " O'Reilly Media, Inc.", 2009.

\bibitem{bladh2004extending}
T.~Bladh, D.~A. Carr, and J.~Scholl.
\newblock Extending tree-maps to three dimensions: A comparative study.
\newblock In {\em Asia-Pacific Conference on Computer Human Interaction}, pp.
  50--59. Springer, 2004.

\bibitem{braun2006using}
V.~Braun and V.~Clarke.
\newblock Using thematic analysis in psychology.
\newblock {\em Qualitative research in psychology}, 3(2):77--101, 2006.

\bibitem{burch2015visualizing}
M.~Burch, T.~Munz, F.~Beck, and D.~Weiskopf.
\newblock Visualizing work processes in software engineering with developer
  rivers.
\newblock In {\em 2015 IEEE 3rd Working Conference on Software Visualization
  (VISSOFT)}, pp. 116--124. IEEE, 2015.

\bibitem{buse2012information}
R.~P. Buse and T.~Zimmermann.
\newblock Information needs for software development analytics.
\newblock In {\em 2012 34th International Conference on Software Engineering
  (ICSE)}, pp. 987--996. IEEE, 2012.

\bibitem{10.5555/2695634}
S.~Chacon and B.~Straub.
\newblock {\em Pro Git}.
\newblock Apress, USA, 2nd ed., 2014.

\bibitem{codoban2015software}
M.~Codoban, S.~S. Ragavan, D.~Dig, and B.~Bailey.
\newblock Software history under the lens: A study on why and how developers
  examine it.
\newblock In {\em 2015 IEEE International Conference on Software Maintenance
  and Evolution (ICSME)}, pp. 1--10. IEEE, 2015.

\bibitem{diehl2007software}
S.~Diehl.
\newblock {\em Software visualization: visualizing the structure, behaviour,
  and evolution of software}.
\newblock Springer Science \& Business Media, 2007.

\bibitem{eick1992seesoft}
S.~Eick, J.~L. Steffen, and E.~E. Sumner~Jr.
\newblock Seesoft-a tool for visualizing line oriented software statistics.
\newblock {\em IEEE Transactions on Software Engineering}, (11):957--968, 1992.

\bibitem{eick2002visualizing}
S.~G. Eick, T.~L. Graves, A.~F. Karr, A.~Mockus, and P.~Schuster.
\newblock Visualizing software changes.
\newblock {\em IEEE Transactions on Software Engineering}, 28(4):396--412,
  2002.

\bibitem{elsen2013visgi}
S.~Elsen.
\newblock Visgi: Visualizing git branches.
\newblock In {\em 2013 First IEEE Working Conference on Software Visualization
  (VISSOFT)}, pp. 1--4. IEEE, 2013.

\bibitem{ens2014chronotwigger}
B.~Ens, D.~Rea, R.~Shpaner, H.~Hemmati, J.~E. Young, and P.~Irani.
\newblock Chronotwigger: A visual analytics tool for understanding source and
  test co-evolution.
\newblock In {\em 2014 Second IEEE Working Conference on Software
  Visualization}, pp. 117--126. IEEE, 2014.

\bibitem{feiner2018repovis}
J.~Feiner and K.~Andrews.
\newblock Repovis: Visual overviews and full-text search in software
  repositories.
\newblock In {\em 2018 IEEE Working Conference on Software Visualization
  (VISSOFT)}, pp. 1--11. IEEE, 2018.

\bibitem{feist2016visualizing}
M.~D. Feist, E.~A. Santos, I.~Watts, and A.~Hindle.
\newblock Visualizing project evolution through abstract syntax tree analysis.
\newblock In {\em 2016 IEEE Working Conference on Software Visualization
  (VISSOFT)}, pp. 11--20. IEEE, 2016.

\bibitem{fluri2007change}
B.~Fluri, M.~Wuersch, M.~PInzger, and H.~Gall.
\newblock Change distilling: Tree differencing for fine-grained source code
  change extraction.
\newblock {\em IEEE Transactions on software engineering}, 33(11):725--743,
  2007.

\bibitem{fritz2010using}
T.~Fritz and G.~C. Murphy.
\newblock Using information fragments to answer the questions developers ask.
\newblock In {\em 2010 ACM/IEEE 32nd International Conference on Software
  Engineering}, vol.~1, pp. 175--184. IEEE, 2010.

\bibitem{froehlich2004unifying}
J.~Froehlich and P.~Dourish.
\newblock Unifying artifacts and activities in a visual tool for distributed
  software development teams.
\newblock In {\em Proceedings. 26th International Conference on Software
  Engineering}, pp. 387--396. IEEE, 2004.

\bibitem{gleicher2017considerations}
M.~Gleicher.
\newblock Considerations for visualizing comparison.
\newblock {\em IEEE transactions on visualization and computer graphics},
  24(1):413--423, 2017.

\bibitem{greene2014conceptcloud}
G.~J. Greene and B.~Fischer.
\newblock Conceptcloud: A tagcloud browser for software archives.
\newblock In {\em Proceedings of the 22nd ACM SIGSOFT International Symposium
  on Foundations of Software Engineering}, pp. 759--762, 2014.

\bibitem{hassan2005top}
A.~E. Hassan and R.~C. Holt.
\newblock The top ten list - dynamic fault prediction.
\newblock In {\em 21st IEEE International Conference on Software Maintenance
  (ICSM'05)}, pp. 263--272. IEEE, 2005.

\bibitem{hattori2011software}
L.~Hattori, M.~D'Ambros, M.~Lanza, and M.~Lungu.
\newblock Software evolution comprehension: Replay to the rescue.
\newblock In {\em 2011 IEEE 19th International Conference on Program
  Comprehension}, pp. 161--170. IEEE, 2011.

\bibitem{hattori2008nature}
L.~P. Hattori and M.~Lanza.
\newblock On the nature of commits.
\newblock In {\em 2008 23rd IEEE/ACM International Conference on Automated
  Software Engineering-Workshops}, pp. 63--71. IEEE, 2008.

\bibitem{heer2010tour}
J.~Heer, M.~Bostock, and V.~Ogievetsky.
\newblock A tour through the visualization zoo.
\newblock {\em Communications of the ACM}, 53(6):59--67, 2010.

\bibitem{heimerl2014word}
F.~Heimerl, S.~Lohmann, S.~Lange, and T.~Ertl.
\newblock Word cloud explorer: Text analytics based on word clouds.
\newblock In {\em 2014 47th Hawaii International Conference on System
  Sciences}, pp. 1833--1842. IEEE, 2014.

\bibitem{hu2014effective}
H.~Hu, H.~Zhang, J.~Xuan, and W.~Sun.
\newblock Effective bug triage based on historical bug-fix information.
\newblock In {\em 2014 IEEE 25th International Symposium on Software
  Reliability Engineering}, pp. 122--132. IEEE, 2014.

\bibitem{johnson1991tree}
B.~Johnson and B.~Shneiderman.
\newblock Tree-maps: A space-filling approach to the visualization of
  hierarchical information structures.
\newblock In {\em Proceedings of the 2nd conference on Visualization'91}, pp.
  284--291. IEEE Computer Society Press, 1991.

\bibitem{kalliamvakou2014promises}
E.~Kalliamvakou, G.~Gousios, K.~Blincoe, L.~Singer, D.~M. German, and
  D.~Damian.
\newblock The promises and perils of mining github.
\newblock In {\em Proceedings of the 11th working conference on mining software
  repositories}, pp. 92--101, 2014.

\bibitem{khan2015interactive}
T.~Khan, H.~Barthel, K.~Amrhein, A.~Ebert, and P.~Liggesmeyer.
\newblock An interactive approach for inspecting software system measurements.
\newblock In {\em IFIP Conference on Human-Computer Interaction}, pp. 1--8.
  Springer, 2015.

\bibitem{kim2007predicting}
S.~Kim, T.~Zimmermann, E.~J. Whitehead~Jr, and A.~Zeller.
\newblock Predicting faults from cached history.
\newblock In {\em 29th International Conference on Software Engineering
  (ICSE'07)}, pp. 489--498. IEEE, 2007.

\bibitem{Kovalenko2018}
V.~Kovalenko, F.~Palomba, and A.~Bacchelli.
\newblock {Mining file histories: Should we consider branches?}
\newblock {\em ASE 2018 - Proceedings of the 33rd ACM/IEEE International
  Conference on Automated Software Engineering}, pp. 202--213, 2018. doi: {{%
10\hspace{.1pt}\discretionary{.}{%
}{.}\hspace{.4pt}1145\discretionary{/}{%
}{/}3238147\hspace{.1pt}\discretionary{.}{%
}{.}\hspace{.4pt}3238169}}


\bibitem{kubelka2019live}
J.~Kubelka, R.~Robbes, and A.~Bergel.
\newblock Live programming and software evolution: questions during a
  programming change task.
\newblock In {\em 2019 IEEE/ACM 27th International Conference on Program
  Comprehension (ICPC)}, pp. 30--41. IEEE, 2019.

\bibitem{latoza2010hard}
T.~D. LaToza and B.~A. Myers.
\newblock Hard-to-answer questions about code.
\newblock In {\em Evaluation and Usability of Programming Languages and Tools},
  pp. 1--6. 2010.

\bibitem{lebeuf2018understanding}
C.~Lebeuf, E.~Voyloshnikova, K.~Herzig, and M.-A. Storey.
\newblock Understanding, debugging, and optimizing distributed software builds:
  A design study.
\newblock In {\em 2018 IEEE International Conference on Software Maintenance
  and Evolution (ICSME)}, pp. 496--507. IEEE, 2018.

\bibitem{lee2013git}
H.~Lee, B.-K. Seo, and E.~Seo.
\newblock A git source repository analysis tool based on a novel
  branch-oriented approach.
\newblock In {\em 2013 International Conference on Information Science and
  Applications (ICISA)}, pp. 1--4. IEEE, 2013.

\bibitem{michaud2016recovering}
H.~M. Michaud, D.~T. Guarnera, M.~L. Collard, and J.~I. Maletic.
\newblock Recovering commit branch of origin from github repositories.
\newblock In {\em 2016 IEEE International Conference on Software Maintenance
  and Evolution (ICSME)}, pp. 290--300. IEEE, 2016.

\bibitem{nagappan2005use}
N.~Nagappan and T.~Ball.
\newblock Use of relative code churn measures to predict system defect density.
\newblock In {\em Proceedings of the 27th international conference on Software
  engineering}, pp. 284--292, 2005.

\bibitem{north2016understanding}
K.~J. North, A.~Sarma, and M.~B. Cohen.
\newblock Understanding git history: A multi-sense view.
\newblock In {\em Proceedings of the 8th International Workshop on Social
  Software Engineering}, pp. 1--7, 2016.

\bibitem{ogawa2009code_swarm}
M.~Ogawa and K.-L. Ma.
\newblock code\_swarm: A design study in organic software visualization.
\newblock {\em IEEE Transactions on Visualization and Computer Graphics},
  15(6):1097--1104, 2009.

\bibitem{perrie2019city}
J.~Perrie, J.~Xie, M.~Nayebi, M.~Fokaefs, K.~Lyons, and E.~Stroulia.
\newblock City on the river: visualizing temporal collaboration.
\newblock In {\em Proceedings of the 29th Annual International Conference on
  Computer Science and Software Engineering}, pp. 82--91, 2019.

\bibitem{porkolab2018codecompass}
Z.~Porkol{\'a}b, T.~Brunner, D.~Krupp, and M.~Csord{\'a}s.
\newblock Codecompass: an open software comprehension framework for industrial
  usage.
\newblock In {\em Proceedings of the 26th Conference on Program Comprehension},
  pp. 361--369, 2018.

\bibitem{rozenberg2016comparing}
D.~Rozenberg, I.~Beschastnikh, F.~Kosmale, V.~Poser, H.~Becker, M.~Palyart, and
  G.~C. Murphy.
\newblock Comparing repositories visually with repograms.
\newblock In {\em Proceedings of the 13th International Conference on Mining
  Software Repositories}, pp. 109--120, 2016.

\bibitem{rufiange2014animatrix}
S.~Rufiange and G.~Melan{\c{c}}on.
\newblock Animatrix: A matrix-based visualization of software evolution.
\newblock In {\em 2014 Second IEEE Working Conference on Software
  Visualization}, pp. 137--146. IEEE, 2014.

\bibitem{sadowski2015developers}
C.~Sadowski, K.~T. Stolee, and S.~Elbaum.
\newblock How developers search for code: a case study.
\newblock In {\em Proceedings of the 2015 10th Joint Meeting on Foundations of
  Software Engineering}, pp. 191--201, 2015.

\bibitem{safwan2019decomposing}
K.~A. Safwan and F.~Servant.
\newblock Decomposing the rationale of code commits: the software developer’s
  perspective.
\newblock In {\em Proceedings of the 2019 27th ACM Joint Meeting on European
  Software Engineering Conference and Symposium on the Foundations of Software
  Engineering}, pp. 397--408, 2019.

\bibitem{shneiderman1996eyes}
B.~Shneiderman.
\newblock The eyes have it: A task by data type taxonomy for information
  visualizations.
\newblock In {\em Proceedings 1996 IEEE symposium on visual languages}, pp.
  336--343. IEEE, 1996.

\bibitem{sillito2006questions}
J.~Sillito, G.~C. Murphy, and K.~De~Volder.
\newblock Questions programmers ask during software evolution tasks.
\newblock In {\em Proceedings of the 14th ACM SIGSOFT international symposium
  on Foundations of software engineering}, pp. 23--34, 2006.

\bibitem{sisman2012incorporating}
B.~Sisman and A.~C. Kak.
\newblock Incorporating version histories in information retrieval based bug
  localization.
\newblock In {\em 2012 9th IEEE Working Conference on Mining Software
  Repositories (MSR)}, pp. 50--59. IEEE, 2012.

\bibitem{stasko2000evaluation}
J.~Stasko, R.~Catrambone, M.~Guzdial, and K.~McDonald.
\newblock An evaluation of space-filling information visualizations for
  depicting hierarchical structures.
\newblock {\em International journal of human-computer studies},
  53(5):663--694, 2000.

\bibitem{stevens2014querying}
R.~Stevens and C.~De~Roover.
\newblock Querying the history of software projects using qwalkeko.
\newblock In {\em 2014 IEEE International Conference on Software Maintenance
  and Evolution}, pp. 585--588. IEEE, 2014.

\bibitem{tao2012software}
Y.~Tao, Y.~Dang, T.~Xie, D.~Zhang, and S.~Kim.
\newblock How do software engineers understand code changes? an exploratory
  study in industry.
\newblock In {\em Proceedings of the ACM SIGSOFT 20th International Symposium
  on the Foundations of Software Engineering}, pp. 1--11, 2012.

\bibitem{tymchuk2016walls}
Y.~Tymchuk, L.~Merino, M.~Ghafari, and O.~Nierstrasz.
\newblock Walls, pillars and beams: A 3d decomposition of quality anomalies.
\newblock In {\em 2016 IEEE Working Conference on Software Visualization
  (VISSOFT)}, pp. 126--135. IEEE, 2016.

\bibitem{ulan2018quality}
M.~Ulan, S.~H{\"o}nel, R.~M. Martins, M.~Ericsson, W.~L{\"o}we, A.~Wingkvist,
  and A.~Kerren.
\newblock Quality models inside out: Interactive visualization of software
  metrics by means of joint probabilities.
\newblock In {\em 2018 IEEE Working Conference on Software Visualization
  (VISSOFT)}, pp. 65--75. IEEE, 2018.

\bibitem{wall2017podium}
E.~Wall, S.~Das, R.~Chawla, B.~Kalidindi, E.~T. Brown, and A.~Endert.
\newblock Podium: Ranking data using mixed-initiative visual analytics.
\newblock {\em IEEE transactions on visualization and computer graphics},
  24(1):288--297, 2017.

\bibitem{wang2019clonecompass}
Y.~Wang, J.~Weatherston, M.-A. Storey, and D.~German.
\newblock Clonecompass: Visualizations for exploring assembly code clone
  ecosystems.
\newblock In {\em 2019 Working Conference on Software Visualization (VISSOFT)},
  pp. 88--98. IEEE.

\bibitem{wessel2019should}
M.~Wessel, I.~Steinmacher, I.~Wiese, and M.~A. Gerosa.
\newblock Should i stale or should i close? an analysis of a bot that closes
  abandoned issues and pull requests.
\newblock In {\em 2019 IEEE/ACM 1st International Workshop on Bots in Software
  Engineering (BotSE)}, pp. 38--42. IEEE, 2019.

\bibitem{wilde2018merge}
E.~Wilde and D.~German.
\newblock Merge-tree: Visualizing the integration of commits into linux.
\newblock {\em Journal of Software: Evolution and Process}, 30(2):e1936, 2018.

\bibitem{yoon2013visualization}
Y.~Yoon, B.~A. Myers, and S.~Koo.
\newblock Visualization of fine-grained code change history.
\newblock In {\em 2013 IEEE Symposium on Visual Languages and Human Centric
  Computing}, pp. 119--126. IEEE, 2013.

\bibitem{zanakis1998multi}
S.~H. Zanakis, A.~Solomon, N.~Wishart, and S.~Dublish.
\newblock Multi-attribute decision making: A simulation comparison of select
  methods.
\newblock {\em European journal of operational research}, 107(3):507--529,
  1998.

\bibitem{zhao2019knowledge}
Y.~Zhao, H.~Wang, L.~Ma, Y.~Liu, L.~Li, and J.~Grundy.
\newblock Knowledge graphing git repositories: A preliminary study.
\newblock In {\em 2019 IEEE 26th International Conference on Software Analysis,
  Evolution and Reengineering (SANER)}, pp. 599--603. IEEE, 2019.

\end{thebibliography}
\end{document}